\documentclass[prl, etal, 
amsmath,amssymb]{revtex4}
\usepackage{graphicx}
\usepackage{dcolumn}
\usepackage{bm}
\usepackage{verbatim}
\usepackage{xcolor} 
\usepackage{braket} 
\usepackage{pstricks-add}  
\usepackage{tikz}
\usepackage[compat=1.0.0]{tikz-feynman}

\usepackage{floatrow}
\usepackage{floatrow}
\usepackage{graphicx}
\usepackage[label font=bf,labelformat=simple]{subfig}
\usepackage{caption}
\usetikzlibrary{graphs}
\usepackage[utf8]{inputenc}
\usepackage[T1]{fontenc}
\usepackage{mathptmx} 
\usepackage{etoolbox}
\usetikzlibrary{decorations.pathmorphing} 
\usetikzlibrary{matrix} 
\usetikzlibrary{arrows} 
\usetikzlibrary{calc} 
\tikzstyle{snakeline} = [decorate, decoration={pre length=0.1cm,
                         post length=0.1cm, snake, amplitude=0.3mm,
                         segment length=1.5mm},thick, magenta, ->]

\tikzset{
    >=stealth', 
    photon/.style={decorate, decoration={snake}, draw},
	provector/.style={decorate, decoration={snake,amplitude=2.5pt}, draw},
	antivector/.style={decorate, decoration={snake,amplitude=-2.5pt}, draw},
    fermion/.style={draw=black, postaction={decorate},
        decoration={markings,mark=at position .55 with {\arrow[draw=black]{>}}}},
    fermionbar/.style={draw=black, postaction={decorate},
        decoration={markings,mark=at position .55 with {\arrow[draw=black]{<}}}},
    fermionnoarrow/.style={draw=black},
    gluon/.style={decorate, draw=black,
        decoration={coil,amplitude=4pt, segment length=5pt}},
    scalar/.style={dashed,draw=black, postaction={decorate},
        decoration={markings,mark=at position .55 with {\arrow[draw=black]{>}}}},
    scalarbar/.style={dashed,draw=black, postaction={decorate},
        decoration={markings,mark=at position .55 with {\arrow[draw=black]{<}}}},
    scalarnoarrow/.style={dashed,draw=black},
    electron/.style={draw=black, postaction={decorate},
        decoration={markings,mark=at position .55 with {\arrow[draw=black]{>}}}},
	bigvector/.style={decorate, decoration={snake,amplitude=4pt}, draw},
}

\usepackage{color}

\usepackage{CJK}

\def\*#1{\mathbf{#1}}

\usepackage{makeidx}

\usetikzlibrary{positioning, shapes.geometric}

\begin{document}


\begin{CJK*}{UTF8}{gbsn}
\title{ Manipulation of Nuclear Isomers with Lasers: Mechanisms and Prospects}

\def\FDU{Key Lab of Nuclear Physics and Ion-beam Application (MoE), Institute of Modern Physics, Fudan University, Shanghai 200433,  China}
\def\SJTU {Department of Physics and Astronomy, Shanghai Jiao Tong University, Shanghai, 200240, China}

\author{Zhiguo Ma} 
\affiliation{\FDU}

\author{Changbo Fu}\email[Corresponding Author: ] {cbfu@fudan.edu.cn} \affiliation{\FDU}
\author{Wanbing He}
\email[Corresponding Author: ] {hewanbing@fudan.edu.cn}
\affiliation{\FDU}
\author{Yugang Ma}
\affiliation{\FDU}

\date{\today}

\begin{abstract}
\end{abstract}
\maketitle

\maketitle

Over one hundred years have passed since the nuclear isomer was first introduced, 
in analogy with chemical isomers to describe long-lived excited nuclear states. 
In 1921, Otto Hahn discovered the first nuclear isomer $^{234m}$Pa.
After that, step by step, it was realized that different types of nuclear isomers exist, including spin isomer, K isomer, seniority isomers, and ``shape and fission'' isomer \cite{walker2020-100yisomer,jain2021NuclearIsomersPrimer}. The spin isomer occurs when the spin change  $\Delta I$ of a transition is very large. The larger $\Delta I$, the lower the electromagnetic transition rates, the longer the  half-lives. 
The K-isomer exists due to the 
significant change in K,
where K is the projection of the total angular momentum on the symmetry axis. 
The seniority isomers arise due to very small transition probability in seniority conserving transitions around semi-magic nuclei, where the seniority, which corresponds to the number of unpaired nucleons, is a reasonably pure quantum number \cite{jain2021NuclearIsomersPrimer}. For a so-called shape isomer,
the inhibition of the decay transition comes from the associated shape changes. It is caused by that a nucleus is trapped in a deformed shape which is its secondary minimum and is hard to decay back to its ground state\cite{jain2021NuclearIsomersPrimer}.

Isomers play very crucial roles in astrophysical nucleosynthesis.
When carrying an astrophysical nucleosynthesis network calculation,
high accurate inputs of nuclear reaction rates  are expected,
and even one reaction rate can dramatically influence the whole
astrophysical evolution network. Typically, when calculating nucleosynthesis rates,
the isotopes are assumed to be in their  ground state,
or their levels are populated according to the thermal-equilibrium
probability distribution.
However, in actual astrophysical circumstances,
those assumptions may not be valid.
After all, the lifetimes of some isomers could be as long as millions of years,
or even longer than the lifetime of our universe.
Therefore,  for these isomers,
they may not equilibrate thermally in an astrophysical event.
In fact, some nuclear isomers,
which play an essential role in nucleosynthesis, should be treated  as special isotopes,
i.e. astromer, as is called in Ref. 
\cite{misch2020AstromersNuclearIsomers}.

Nuclear isomers  may also be used to make nuclear batteries.
Nuclear batteries usually have a much higher energy density than chemical ones. $^{178m}$Hf has a lifetime of 31 years, and is considered as an  good energy source candidate for deep space exploration.
Furthermore, to obtain a high accuracy clock, nuclear isomers, which generally have much more stable decay transitions ($\Delta E/E$) than atomic ones,
are expected to be used as standards of the next-generation clocks.

With the development of modern laser technologies,
a laser intensity of over $10^{23}$~W/cm$^2$ has been achieved. In a laser beam with an electrical field of $10^{14}$~V/m, in a half-circle of the oscillation,
an electron can acquire energy as high as 100~MeV.
Many laser-plasma-based accelerators are proposed and realized,
including the laser wakefield accelerator,
the plasma beat wave accelerator,
and the self-modulated laser wakefield accelerator etc.\cite{Esarey2009} It has been demonstrated that many classical known nuclear reactions
can be carried out with  high-intensity lasers (HILs),
such as photo-induced fission,
proton- and ion-induced reactions,
and nuclear fusions, etc.\cite{2006Lasers_and_Nuclei}

Instead of being used as accelerators in analogy with traditional ones,
HILs provide unique environments at the extreme for many scientific studies \cite{Fu2022.MRE.024201}. 
Today, HILs can provide
electrical fields with the order of $10^{14}$~V/m,
magnetic fields with the order of $10^8$~T,
pressure with the order of $10^{16}$~P,
currents with the order of $10^{5}$~A,
and temperature with the order of $10^{5}$ to $10^{10}$~K.
With help from HILs,
those extreme features are now available in
earth-based labs for nuclear studies, including those related to nuclear isomers.
Taking advantage of 
very short beam pulses,   
extremely high electric and magnetic fields,
and high beam intensities etc., 
one may have more nuclear isomers manipulating strategies compared with traditional accelerators. 
In the following, the possible mechanisms of manipulating nuclear isomers will be discussed.

In general, nuclear transitions are almost independent of the atomic environment in which the nucleus is.
This is the result that nuclear transitions are normally dominated by the strong force.
The other three fundamental forces, electromagnetic, weak, and gravitational,
are only playing unimportant roles generally.

However, under some special circumstances, things may go differently.
For example, for a $\beta$ decay that involves the weak force,
whether 
an atomic electron
around a $\beta$-decay nuclei
is available
or not may dramatically affect the decay rate.
In fact, as an example, a neutral $^7$Be atom has a lifetime of 54 days, while $^7$Be$^{4+}$ is stable.
A $^7$Be  nucleus would capture an electron (typically from the lowest K-shell) and then decay to $^7$Li. If there is no electron around, the electron capture process is blocked.

Like the weak force, the electromagnetic force may play an essential role in nuclear transitions under some particular circumstances like isomer decays. 
It is worth pointing out that there is a significant difference between nuclear isomer states and radioactive ground states (g.s.).
A nuclear isomer has the possibility of emitting an $\alpha$, proton, neutron,  $e^{\pm}$, or
any other particles, as well as a photon,
while a radioactive isotope at its g.s. may emit everything listed above but a photon.
This means that it is possible to manipulate isomers electromagnetically with high efficiency\cite{walker2020-100yisomer}.
Obviously, compared with the other three fundamental forces,
the weak, strong, and gravitational forces,
the electromagnetic force is much easier to be applied with current techniques.

Manipulating a nuclear isomer  electromagnetically has another advantage, which is that a much higher cross section is possible.
After all, the electromagnetic force is a long-range force.
Furthermore, with an electron cloud around an isomeric nucleus,
the resonance effect can boost to a higher cross section.

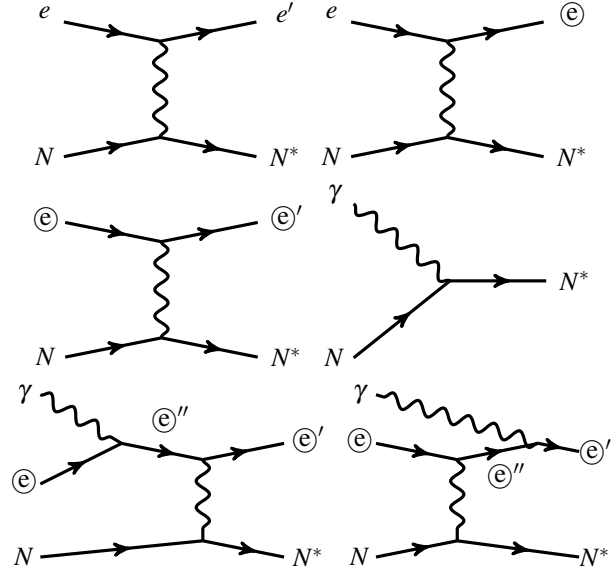
\begin{figure}[htb]
\centering
\begin{tikzpicture}[line width=1.2 pt, scale=1.28]
   \pgfmathsetmacro{\L}{1.0}
	\draw[fermion]    (0   ,0)      --( 1*\L, 0.2*\L);
	\draw[fermionbar] (0   ,0)      --(-1*\L, 0.2*\L); 
	\draw[fermionbar] (1*\L,-1.2*\L)--(    0,-1.0*\L);	
	\draw[fermionbar] (0   ,-1*\L)  --(-1*\L,-1.2*\L);
	\draw[photon]     (0   ,0)      --(    0,-1.0*\L);
	
	\node at (-1.2*\L, 0.3*\L) {\normalsize $e$}; 
	\node at (-1.2*\L,-1.2*\L) {\normalsize $N$}; 
	\node at ( 1.3*\L, 0.3*\L) {\normalsize $e'$}; 
	\node at ( 1.3*\L,-1.2*\L) {\normalsize $N^{*}$}; 
\end{tikzpicture}
\begin{tikzpicture}[line width=1.2 pt, scale=1.28]
   \pgfmathsetmacro{\L}{1.0}
	\draw[fermion]    (0   ,0)      --( 1*\L, 0.2*\L);
	\draw[fermionbar] (0   ,0)      --(-1*\L, 0.2*\L); 
	\draw[fermionbar] (1*\L,-1.2*\L)--(    0,-1.0*\L);	
	\draw[fermionbar] (0   ,-1*\L)  --(-1*\L,-1.2*\L);
	\draw[photon]     (0   ,0)      --(    0,-1.0*\L);
	
	\node at (-1.2*\L, 0.3*\L) {\normalsize $e$}; 
	\node at (-1.2*\L,-1.2*\L) {\normalsize $N$}; 
	\node at ( 1.3*\L, 0.25*\L) {\normalsize $\textcircled{e}$}; 
	\node at ( 1.3*\L,-1.2*\L) {\normalsize $N^{*}$}; 
\end{tikzpicture}

\begin{tikzpicture}[line width=1.2 pt, scale=1.28]
   \pgfmathsetmacro{\L}{1.0}
	\draw[fermion]    (0   ,0)      --( 1*\L, 0.2*\L);
	\draw[fermionbar] (0   ,0)      --(-1*\L, 0.2*\L); 
	\draw[fermionbar] (1*\L,-1.2*\L)--(    0,-1.0*\L);	
	\draw[fermionbar] (0   ,-1*\L)  --(-1*\L,-1.2*\L);
	\draw[photon]     (0   ,0)      --(    0,-1.0*\L);
	
	\node at (-1.2*\L, 0.2*\L) {\normalsize \textcircled{e}}; 
	\node at (-1.2*\L,-1.2*\L) {\normalsize $N$}; 
	\node at ( 1.3*\L, 0.25*\L) {\normalsize \textcircled{e}$'$}; 
	\node at ( 1.3*\L,-1.2*\L) {\normalsize $N^{*}$}; 
\end{tikzpicture}
\begin{tikzpicture}[line width=1.2 pt, scale=1.28]
   \pgfmathsetmacro{\L}{1.0}
   \pgfmathsetmacro{\H}{-0.5}
	\draw[photon]     (0   ,0+\H)      --(-1*\L, 0.8*\L+\H); 
	\draw[fermionbar] (0   ,0+\H)      --(-1*\L,-0.8*\L+\H);
	\draw[fermionbar] (1*\L,0*\L+\H)   --(    0,0+\H);	
	
	\node at (-1.2*\L, 0.4*\L) {\normalsize $\gamma$}; 
	\node at (-1.2*\L,-1.3*\L) {\normalsize $N$}; 
	\node at ( 1.3*\L,-0.5*\L) {\normalsize $N^{*}$}; 
\end{tikzpicture}

\begin{tikzpicture}[line width=1.2 pt, scale=1.08]
   \pgfmathsetmacro{\L}{1.0}
   \pgfmathsetmacro{\H}{-0.5}
	\draw[fermion]    (0   ,0)      --( 1*\L, 0.2*\L);
	\draw[fermionbar] (0   ,0)      --(-1*\L, 0.2*\L); 
	\draw[fermionbar] (1*\L,-1.2*\L)--(    0,-1.0*\L);	
	\draw[fermionbar] (0   ,-1*\L)  --(-2*\L,-1.2*\L);
	\draw[photon]     (0   ,0)      --(    0,-1.0*\L);
	
	\draw[photon]    (-2*\L,  0.8*\L)     --(-1*\L, 0.2*\L); 
	\draw[fermion]   (-2*\L,-0.3*\L)     --(-1*\L, 0.2*\L);

	\node at (-0.4*\L, 0.5*\L) {\normalsize \textcircled{e}$''$};
	\node at (-2.2*\L,-1.2*\L) {\normalsize $N$};
	\node at ( 1.3*\L, 0.25*\L) {\normalsize \textcircled{e}$'$}; 
	\node at ( 1.3*\L,-1.2*\L) {\normalsize $N^{*}$}; 
	\node at ( -2.2*\L,-0.3*\L) {\normalsize  $\textcircled{e}$}; 
	\node at ( -2.2*\L, 0.8*\L) {\normalsize $\gamma$}; 
\end{tikzpicture}
\begin{tikzpicture}[line width=1.2 pt, scale=1.08] \label{fig.eb}
   \pgfmathsetmacro{\L}{1.0}
   \pgfmathsetmacro{\H}{-0.5}
	\draw[fermion]    (0   ,0)      --( 1*\L, 0.2*\L);
	\draw[fermionbar] (0   ,0)      --(-1*\L, 0.2*\L); 
	\draw[fermionbar] (1.5*\L,-1.2*\L)--(    0,-1.0*\L);	
	\draw[fermionbar] (0   ,-1*\L)  --(-1*\L,-1.2*\L);
	\draw[photon]     (0   ,0)      --(-0*\L,-1.0*\L);
	
	\draw[photon]    ( 1.0*\L,0.2*\L)  -- (-1*\L,  0.8*\L); 
	\draw[fermion]   ( 1  *\L,0.2*\L)   --  (1.5*\L,0.1*\L);
        
	\node at ( 0.6*\L,-0.2*\L) {\normalsize \textcircled{e}$''$};
	\node at (-1.2*\L,-1.2*\L) {\normalsize $N$};
	\node at ( 1.7*\L, 0.1*\L) {\normalsize \textcircled{e}$'$}; 
	\node at ( 1.7*\L,-1.2*\L) {\normalsize $N^{*}$}; 
	\node at (-1.2*\L, 0.2*\L) {\normalsize {$\textcircled{e}$}}; 
	\node at (-1.2*\L, 0.8*\L) {\normalsize $\gamma$}; 
\end{tikzpicture}

\caption{
  The Feynman diagrams of the possible mechanisms to manipulate nuclear isomers,
  (a) CE, (b) NEEC, (c) NEET, (d) PE, and (e) EB.  
The  ``\textcircled{$e$}'' indicates that the electron is at
an atomically bounded state, while  ``$e$'' at a continuum state.  
$N$ represents the nucleus is on its g.s.,
while $N^{*}$ on an excited state.
}
\label{fig.feynman}
\end{figure}

A nuclear isomer may be manipulated through the mechanisms of Coulomb excitation (CE),
nuclear excitation by electron capture (NEEC),
nuclear excitation by electron Transition (NEET),
direct photon excitation (PE),
and electron bridge (EB), etc. \cite{Harston-PRC59(1999)2462} 
In a typical laser-induced plasma,
all these mechanisms are possible,
though some of them have not been confirmed experimentally yet.
The CE and PE are relatively well-known.
The NEET has been observed in $^{197}$Au  \cite{NEET-PRL2000-Au197}.
It was reported that the NEEC transition was observed in $^{93}$Mo
\cite{Chiara2018-Nature554-216},
although more evidence is needed to confirm that \cite{93mMo-PRL2019}.
The Feynman diagrams of different mechanisms of manipulating isomers are shown in Fig.\ref{fig.feynman}, 
and some of them, CE, NEEC, NEET, and EB, will be discussed in detail in the following. 

The nuclear CE is an inelastic scattering process
in which a nucleus is excited due to the electromagnetic field when a
charged particle passes by 
\cite{Alder:1956im}.
This process can happen at an energy below the Coulomb
barrier since only the Coulomb force, and not the short-range nuclear forces, is involved in the process.

For electric and magnetic  excitation, the cross sections in units of barns can be written as\cite{Alder:1956im}
\begin{eqnarray}
\label{eq:1}
  \sigma_{E\lambda}&=&c_{E\lambda} E_{MeV}^{\lambda- 2}(E_{MeV}
 -\Delta E_{MeV}')^{\lambda-1} B(E\lambda)
                     f_{E\lambda(\eta_i,\xi)} , \\
\sigma_{M\lambda}&=&c_{M\lambda} E_{MeV}^{\lambda-
                     \frac{3}{2}}(E_{MeV}
-\Delta E_{MeV}')^{\lambda-\frac{1}{2}} B(M\lambda)
                     f_{M\lambda(\eta_i,\xi)},
\end{eqnarray}
with
$c_{E\lambda}= \frac{Z_1^2 A_1}{40.03}
\left[0.07199\left(1+\frac{A_1}{A_2}\right)Z_1Z_2\right]^{-2\lambda+2}$, and 
$c_{M\lambda}= 5.888\times 10^{-9}Z_1^{2}\left[0.07199\left(1+\frac{A_1}{A_2}\right)Z_1Z_2\right]^{-2\lambda+2}$, 
where $E_{MeV}$ is the projectile's energy in unit MeV,
$\Delta E_{MeV}'=(1 +A_1/A_2)·\Delta E_{MeV}$ and 
$\Delta E_{MeV}$
represents the excitation energy,
$A_1$($A_2$) is the mass of the projectile (target),
$B(E\lambda)$ [$B(M\lambda)$] represents the  reduced transition
probability associated with a radioactive transition of multipole order $E\lambda$  [$M\lambda$] in units of $e^2\cdot (10^{-24} cm^{2})^{\lambda}$ { $[(e\hbar/2Mc)^2\cdot (10^{-24} cm^{2})^{\lambda-1}]$}, the $E\lambda$ represents the electrical transition,
and $M\lambda$ represents the magnetic transition,
$\lambda$ is the order of electric (magnetic) multipole component,
and $f_{E\lambda}(\eta_i,\xi)$ is the f-function described in Ref.\cite{Alder:1956im}.

In laser-induced plasma, a nucleus can be excited by other nuclei,
as well  as by high energy electrons accelerated by laser fields,
through the CE mechanism.

First proposed by Goldanskii, the NEEC is a nuclear excitation process after a free electron being
captured by the atom, which is
$N+e\rightarrow N^*+\textcircled{e}$,
where  the original electron $e$ is at a continuum state,
and the final electron is marked as ``$\textcircled{e}$'' to indicate that it is atomically  bound.

The NEEC rate in a plasma with electron temperature $T_e$ can be
written as \cite{PRC:2004:Gosselin}
\begin{eqnarray}\label{eq.NEEC}
\lambda^{NEEC}&=&\frac{\alpha(T_e)\ln 2}{T^{\gamma}_{I_f\rightarrow
    I_i}}\frac{2I_f+1}{2I_i+1}
                  f_{FD}(E_r)[1-f_{FD}(E_b)] \nonumber 
\times \frac{1}{2}
\left[
 {\rm Erf}\left(\frac{E_r}{\epsilon\sqrt{2}}\right)
-{\rm Erf}\left(\frac{E_b}{\epsilon\sqrt{2}}\right)
\right]
,
\end{eqnarray}
where 
$I_i$ and $I_f$ are initial and final nuclear spin, respectively,
$\Gamma^{tot}$ is the total transition width,
$T^{\gamma}_{I_i\rightarrow I_f}$ is the $\gamma$-transition width of $I_i\rightarrow I_f$. $f_{FD}$ is the Fermi-Dirac statistics function, $E_{b}$ is the binding energy,
and $E_r$ is the resonance energy,
${\rm Erf}(x)\equiv\int_0^xe^{-t^2}{\rm d}t$ is the error function, and $\epsilon$ is the dispersion of the electronic transition energy of the real configuration around the average atom value.
The $\alpha(T_e)$ is the internal conversion coefficient (ICC),
which depends on the electron temperature $T_{e}$.
ICC $\alpha$ is defined as 
$\alpha=\sum_i\alpha_i=\sum_i\frac{\kappa_i}{\kappa_\gamma}$,
where $\kappa_i$ is the branching ratio of the radioactive nucleus  emitting an electron from $i$-th atomic shell, and $\kappa_{\gamma}$ is the branching ratio of the nucleus emitting an photon $\gamma$.

NEEC is the inverse process of a well-known nuclear decay, internal
conversion.
In plasma, according to the detailed balance principle, the ratio of the NEEC excitation rate
to the internal conversion decay rate is written as,
\begin{equation}
\label{eq:3}
\frac{\lambda_e^{NEEC}}{\lambda_d^{IC}}=\frac{2I_f+1}{2I_i+1}
e^{-\Delta E/k_BT_e}.
\end{equation}
The same as excited by a free electron.
It is clear that the ratio of the NEEC rate and the photon excitation rate is proportional to the internal conversion coefficient
$\alpha$. 
For a nuclear isomer that has $\alpha$  much larger than 1, $\alpha \gg 1$, 
it is much easier to excite it by electrons than by photons.
Therefore, as the inverse processes, 
for $\alpha\gg 1$ nuclei, 
$N+e\rightarrow N^*+\textcircled{e}$  is a much more efficient way to excite  the nuclei $N$ than  
$\gamma+N\rightarrow N^*$.


The NEET process was first proposed by Morita in 1976. 
The process is  \cite{NEET-PRL2000-Au197,NEET-Os189-PRC2000}
$N+\textcircled{e}^{*}\rightarrow N^*+\textcircled{e}$.
In weak coupling limit ($\kappa\rightarrow 0$), the NEET cross section can be written as
\cite{NEET-Os189-PRC2000} \begin{equation}\label{eq.NEET}
\sigma_{NEET}^{\kappa\rightarrow 0}=\frac{\Gamma_i\Gamma_f}{\Gamma_i}
\frac{\kappa^2}{(E_i-E_f)^2+\left(\frac{\Gamma^{tot}}{2}\right)^2},
\end{equation}
where $\kappa=\braket{f|i}$,
the $i$ and $f$ represent the initial and final states respectively,
$\Gamma$ is the transition width,
and $E$ is the binding energy.

In a laser-plasma environment where a lot of energetic electrons are around, 
NEEC and NEET may dominate the processes of generating nuclear isomers.

The EB process is
$N+\gamma+\textcircled{e}^{*}\rightarrow N^*+\textcircled{e}$.
It is also known as ``laser assistant NEET'' in some literature.
{ One can find where the name comes from by comparing the Feynman diagrams in Fig.\ref{fig.feynman}(c) with the (e),
In the EB process, an extra photon is needed.}
Because of the energy conservation,
$    E_{N_0}+E_{\gamma}+E_{e_i}=E_{N*}+E_{e_f}$,
one has
$    \Delta E_N\equiv E_{N*}-E_{N_0}=E_{\gamma}-(E_{e_f}-E_{e_i})=E_{\gamma}-\Delta E_e$.
By choosing a proper $E_{\gamma}$ and  $\Delta E_e\equiv E_{e_f}-E_{e_i} $,
one has a convenient experimental way to match the nuclear excited energy $\Delta E_N$.

As shown in Fig.\ref{fig.eb}, 
there are two possible ways of coupling between the incoming photon and the electron: the $e$-$\gamma$ coupling vertex is before or after the coupling of the electron with the nucleus.  
The cross section of the EB process can be written as \cite{Borisyuk2019-PRC.100.044306},
\begin{equation}
\label{eq:4}
\sigma_{EB}(\omega_L)=g
\frac{\Gamma^{tot}\Gamma^{IC}(\omega_{IC};f)}{(\omega_L-\omega_L^{res})^2+(\Gamma^{tot}/2)^2}\sigma_{atom}(\omega_L),
\end{equation}
where $\Gamma^{tot}(\omega_{IC};f)$ is the partial width
(probability) of the internal conversion from the $\ket{f}$ shell
of the atom at the virtual photon energy
$\omega_{IC}=\omega_L+E_i-E_f$,
$\sigma_{atom}$ is the photoionization cross section , $\Gamma^{tot}$ is the total width of the nuclear level,
$\omega_L^{res}=E_f-E_i+\omega_N$ is the resonance energy of the laser
photons, and
$g=\frac{2J_f+1}{2J_c+1}
\frac{2I_f+1}{2I_i+1}
\frac{1}{(2J_c+1)^2}$.


At resonance, $\omega_L=\omega_L^{res}$, the formula above is simplified to be,
\begin{equation}
\label{eq:7}
\sigma_{EB}(\omega_L^{res},i\rightarrow f)
\propto \Gamma^{IC}\sigma_{atom}(\omega_L^{res}). 
\end{equation}
One can find that the cross section is proportional to the
photoionization cross section $\sigma_{atom}$, which is a significant enhancing factor.

{
As shown above, 
with help of lasers, especially HILs,
nuclear isomers can be manipulated through the mechanisms of CE, NEEC, NEET, PE,  EB, etc. 
However, from the experimental point of view, there is a difficulty to be overcome.
When HILs hit on targets, 
strong background radiation can be induced, which makes it difficult to distinguish these mechanisms from the background noises.
Recently, by using a tabletop hundred-TW laser system,
pumping $^{83}$Kr nuclei to their nuclear isomeric states has been
achieved \cite{Feng.2022.PRL.128.052501}.
}

{Recently, $^{83m}$Kr Induced by a fs Laser has been observed for the  first time.}
In this experiment\cite{Feng.2022.PRL.128.052501},
natural Kr gas, with 11.5\% of $^{83}$Kr isotope,
was blown out from a jet with a backing pressure up to 7  MPa.
During the adiabatic expansion process,
the Kr gas jet was cooled down and frozen to be nano-clusters.
The laser then shot at the clusters,
and the quivering electrons induced by laser fields can excite
$^{83}$Kr nuclei to their excited states.
The reaction products were then collected by a liquid-nitrogen cooled trap.
The decay of $^{83m}$Kr$_{2}$ (E=41.6 keV, $T_{1/2}=1.83$ h) isomers was recorded by NaI detectors  close to the trap.
A  peak production of $2.34 \times 10^{15}$ particles/s has been
observed.

Taking advantage of fast and powerful fs lasers,
this  universal production method can be widely used for pumping
isotopes with excited state lifetimes down to picoseconds,
and could benefit fields like nuclear transition mechanisms and nuclear $\gamma$-ray lasers.

In this experiment, the observed products, the 2nd excited state $^{83m}$Kr$_{2}$ (E=41.6 keV,
$T_{1/2}=1.83$ h), could be explained by CE.
It is still an open question
whether the 1st excited state $^{83m}$Kr$_{1}$ (E=9.4 keV,
$T_{1/2}=159$ ns) may also be populated through the process like NEEC,
NEET, and PE, etc. 

{
Worldwidely, many HILs have been built or are under construction.
They can provide extreme environments with plasma temperature over keV, which covers many isomers' excitation energy ranges. 
It is expected that these HIL facilities would provide a new and convenient platform to manipulate nuclear isomers. 
Following the first successfully discovering laser-induced isomer of $^{83}Kr$,  we believe that more mechanisms, including long expected NEEC etc., of manipulating isomers would be confirmed in the near future. 
}

This work is supported by
the National Nature Science Foundation of China (under grant no. 11875191)
and the Strategic Priority Research Program of the Chinese Academy of Sciences (grant no. XDB16).

\end{CJK*}

\end{document}